\documentclass{aa}
\usepackage{epsfig,graphics}
\usepackage[dvips]{rotating}
\usepackage{latexsym}
\usepackage{graphicx}
\usepackage{subfigure}
\usepackage{amsmath,multirow,amsfonts}

\newcommand{\cd}{d$^{-1}$}

\newcommand{\kms}{kms$^{-1}$}

\newcommand{\vsini}{$v\sin{i}$}
\newcommand{\logg}{\ensuremath{\log g}}

\begin{document}

\title{A new method for the spectroscopic identification of stellar non-radial pulsation modes}
\subtitle{I. The method and numerical tests}

\author{W. Zima$^1$}

\offprints{zima@ster.kuleuven.be}

\institute
{
$^1$ Institut f\"ur Astronomie der Universit\"at Wien, 
T\"urkenschanzstr. 17, A - 1180 Wien, Austria}

\date{Received / Accepted }
\abstract{}
{We present the Fourier parameter fit method, a new method for spectroscopically identifying stellar radial and non-radial pulsation modes based on the high-resolution time-series spectroscopy of absorption-line profiles. In contrast to previous methods this one permits a quantification of the statistical significance of the computed solutions. The application of genetic algorithms in seeking solutions makes it possible to search through a large parameter space.} 
{The mode identification is carried out by minimizing $\chi^2$, using the observed amplitude and phase across the line profile and their modeled counterparts. Computations of the theoretical line profiles are based on a stellar displacement field, which is described as superposition of spherical harmonics and that includes the first order effects of the Coriolis force.} 
{We made numerical tests of the method on a grid of different mono- and multi-mode models for $0 \le \ell \le 4$ in order to explore its capabilities and limitations. Our results show that whereas the azimuthal order $m$ can be unambiguously identified for low-order modes, the error of $\ell$ is in the range of $\pm 1$. The value of $m$ can be determined  with higher precision than with other spectroscopic mode identification methods. Improved values for the inclination can be obtained from the analysis of non-axisymmetric pulsation modes. The new method is ideally suited to intermediatley rotating $\delta$ Scuti and $\beta$ Cephei stars.} 
{}


\maketitle

\markboth{Zima : 
A new method for the spectroscopic identification of stellar non-radial pulsation modes}
{Zima : 
A new method for the spectroscopic identification of stellar non-radial pulsation modes}

\section{Introduction}
Asteroseismology relies on a detailed comparison between the set of frequencies observed in a pulsating star and those predicted by theory. To do this successfully, we need to know the quantum numbers associated with each observed pulsation mode, and thus we need {\it mode identifications} for as many frequencies as possible. The combined approach of spectroscopic identification methods, multicolor photometry, and frequency pattern recognition can identify both the harmonic degree $\ell$ and the azimuthal order $m$ of a pulsation mode. The parameters of a stellar seismic model, such as the mass, the luminosity, the rotation rate, the effective temperature, the metallicity, and the age, can be tuned to obtain a fit between the theoretical and observed frequencies. In most cases the range of possible parameters is very large. By only considering the models that are consistent with mode identification, one is able to put a strong constraint on the internal structure of the star. For an extensive overview paper about photometric mode identification methods we refer the reader to Garrido (2000), and to Balona \& Evers (1999) for a paper about the state-of-the-art approach of photometric mode identification.

Profile variations of absorption lines formed in the photosphere can be used to study the pulsational behavior of a star. The line-profile variations (LPV) are caused by the Doppler shift of absorption lines originating from different parts of the surface. 
High-resolution time-series spectroscopy is a powerful tool for identifying the $\ell$ and $m$ values of a non-radially oscillating star by analyzing its LPV. In addition, these techniques allow determination of the inclination of the pulsation/rotation axis, the pulsational temperature variations, and the intrinsic oscillation amplitude. If photometric amplitudes in different passbands are also known, it is possible to calculate the ratio of flux perturbation to radial displacement, which is very sensitive to convection, thereby permitting selection between different convection models (Daszynska-Daszkiewicz et al. 2003).

In the past decade, considerable effort has been expended to develop and test {\it reliable} spectroscopic mode identification methods. Basically, two kinds of approaches have been developed: one based on quantities integrated across the line profile and one based on the intensity variations within each wavelength bin. The first approach is utilized in the moment method (Balona 1986, Aerts 1996, Briquet \& Aerts 2003) and is applicable to the identification of low-degree modes ($\ell \le 3$). Different variants exist of the second approach, which is based on the Doppler image principle and requires sufficient rotational broadening of the profile. These include the intensity-period search (IPS) method (Schrijvers et al. 1997), where $\ell$ and $m$ are determined from the amplitude and phase distribution across the line profile by applying empirically derived relations, and the pixel-by-pixel method (PPM) described by Mantegazza (2000).

These spectroscopic mode identification methods have one thing in common: no statistical criterion that can quantify the significance of the obtained solutions can be calculated yet. For instance, the moment method (MM) quantifies the goodness-of-fit in elaborately weighted discriminants, which are not related to the observational uncertainties. Therefore it is often not possible to select between different solutions if their discriminants have similar values. De Ridder et al. (2003) report pessimistic prospects for defining such error estimates in the future because of the complex nature of the moments. 

We have developed a spectroscopic mode identification method, the Fourier parameter fit (FPF) method, which relies on a fit of the observational and modeled Fourier parameters across the line profile. For every detected pulsation frequency, the zero point, amplitude, and phase are computed for every wavelength bin across the profile by a multi-periodic least-square fit. The observational errors of these parameters can be calculated analytically from the least-square fit. This enables us to carry out a classical chi-square ($\chi^2$)-test and to derive the statistical significance of our fits.

The paper is constructed in the following way: we first provide the formalism of the displacement field used for the computation of the LPV. The FPF method is presented in the next section, followed by the description of numerical experiments applied for testing the method.

\section{Modeling of the line profile variations}
We assume that the displacement field of a pulsating star can be described by a superposition of spherical harmonics.  Our description of the Lagrangian displacement field is valid in the limit of slow rotation taking the effects of the Coriolis force to the first order into account (Schrijvers et al. 1997). Since deviations from spherical symmetry due to centrifugal forces are ignored, our formalism is reliable only for pulsation modes where the ratio of the rotation to the oscillation frequency $\Omega/\omega < 0.5$ (Aerts \& Waelkens 1993). This limitation excludes realistic modeling of rapidly rotating stars and low-frequency $g$-modes. For higher frequency $p$-modes, such as observed in many $\delta$ Scuti and $\beta$ Cephei stars, the given criterion is fulfilled and a correct treatment is provided. 

The intrinsic line profile is assumed to be a Gaussian. This is a good approximation for shallow lines where the rotational broadening dominates other line broadening mechanisms. A distorted profile is computed from a weighted summation of Doppler shifted profiles over the visible stellar surface. Additionally, we take into account a parameterized variable equivalent width due to temperature and brightness variations across the stellar surface. 

We implemented the calculation of the pulsationally distorted absorption line profiles in a C-program code called {\sc lips}, which can be obtained from the author upon request. The code is also capable to computing a displacement field whose symmetry axis is not aligned with the rotation axis, as observed in roAp stars (Kurtz 1982). Flux variations in different filters can be computed by considering flux derivatives calculated from static atmospheres. A detailed description of the formalisms used for constructing the line profiles can be found in Appendix A.

\section{The $\chi^2_\nu$ of the FPF method}

Since the behavior of the LPV can be very complex due to the multi-periodicity of a star, the contribution of every single pulsation mode to the total profile variations must be separated. Thereby, it is possible to apply techniques developed for one-dimensional time-series data, if the intensity variations of every pixel or wavelength bin across the profile are analyzed separately. 

The search for pulsation frequencies can be carried out by applying a Fourier analysis for every pixel of the profile. For all detected frequencies it is possible, by means of least-square fitting, to determine the zero point $Z_\lambda$, the amplitude $A_\lambda$, and the phase $\phi_\lambda$ across the line. We can thereby quantify the LPV caused by a pulsation mode by its Fourier parameters, which are a function of wavelength or Doppler velocity. 




We directly fit the observed Fourier parameters with theoretical values by applying an $\chi^2$-test. The main differences to previous, similar mode identification methods are the following: 
\begin{itemize}
\item We utilize {\it all} available information on the Fourier parameters, zero point, amplitude, and phase across the line profile. 
\item The calculation of $\chi^2$ enables us to {\it quantify the significance} of our fits.
\item The optimization is carried out with {\it genetic algorithms}, which permit the detection of local minima in a large multi-parameter space in much shorter computational time than a grid allows.
\end{itemize}

We calculate the theoretical line profiles with the profile generation code {\sc lips} described in the first part of this paper. The theoretical Fourier parameters are computed from a least-square fit of typically ten synthetic mono-mode line profiles evenly sampled over one pulsation cycle. As for the observational Fourier parameters, here we also fix the frequency during the least-square fit of the wavelength bins and only derive $Z_\lambda$, $A_\lambda$, and $\phi_\lambda$.

The observational variance $\sigma^2_\lambda$ of the intensity of every pixel across the profile must be derived very carefully, since the value of $\chi^2$ is very sensitive to this variance. Generally, $\sigma^2_\lambda$ is not constant along the profile, but increases towards its center due to the lower signal present there. The variance can be derived directly from the error-matrix of the multi-period least-square fit. 

Our approach is the following: we first determine the pulsationally independent stellar parameters \vsini, equivalent width $W_0$, and intrinsic line width $\sigma$ by fitting the observed zero point profile $Z_\lambda$. We assume $Z_\lambda$ to be independent of the pulsation, which is true for a low ratio of the radial velocity amplitude to \vsini. During the optimization process, we set the values of these parameters as variable in a narrow range set by their derived uncertainties. 

The parameters of the pulsation modes are derived only from the amplitude and phase across the line. The reduced $\chi^2_\nu$ is calculated from complex amplitudes in order to combine amplitude and phase information in the following way 
\begin{equation}
\chi^2_\nu=\frac{1}{2n_\lambda-m}\sum^{n_\lambda}_{i=1}\Biggl[\frac{(A^o_{R,i}-A^t_{R,i})^2}{\sigma^2_{R,i}}+\frac{(A^o_{I,i}-A^t_{I,i})^2}{\sigma^2_{I,i}}\Biggr]
\label{eq:chisq}
\end{equation}
where $n_\lambda$ is the number of pixels across the profile, $m$ the number of free parameters, $A^o$ and $A^t$ denote observationally and theoretically determined values, respectively, $A_R=A_\lambda \cos\phi_\lambda$ and $A_I=A_\lambda \sin\phi_\lambda$ are the real and imaginary parts of the complex amplitude, and $\sigma$ is the observational error. Equation~\ref{eq:chisq} can be easily modified if the observed amplitude and phase from photometric passbands is included for the calculation of $\chi^2_\nu$.

Since the amplitude and phase of a given wavelength bin are treated as independent variables, the variances are calculated from
\begin{equation}\sigma^2_{R,\lambda}=\sigma(A_\lambda)^2 \cos^2 \phi_\lambda +\sigma(\phi_\lambda)^2 A^2_\lambda \sin^2 \phi_\lambda \label{eq:varr}\end{equation}
\begin{equation}\sigma^2_{I,\lambda}=\sigma(A_\lambda)^2 \sin^2 \phi_\lambda +\sigma(\phi_\lambda)^2 A^2_\lambda \cos^2 \phi_\lambda. \label{eq:vari}\end{equation}


In the case of a multi-periodic star, it is possible to optimize the sum of the $\chi^2_\nu$ values of all present modes by simultaneously taking common values for the pulsationally independent parameters inclination, \vsini, $\sigma$, and $W$. By doing so, the stellar inclination angle, and henceforth also the modes' intrinsic pulsation amplitudes, can be reproduced better, since we take advantage of the fact that all observed modes are seen at the same aspect angle. Such an improvement in the identification of modes in multi-periodic stars has also been obtained by Briquet \& Aerts (2003) for the case of the MM.

We carry out the search for the best-fit model by means of a genetic optimization algorithm (Nissen 1997) by allowing the exploration of a large parameter space in much shorter computing time than can be done by grid calculations. The basis of a genetic code is a stochastically selected set of models, which evolves in time towards better models by favoring those having a higher fitness (lower $\chi^2_\nu$). The strength of genetic algorithms is especially their ability to explore many local minima simultaneously and consequently to locate the absolute minimum of a complex multi-parameter optimization problem.

\section{Testing the FPF method} \label{sec:testing}
We tested the functionality of the FPF method in a classical way via numerical experiments by fitting simulated observations of mono-mode pulsations. We especially examined to what precision different parameters of the input modes can be determined, with an emphasis on $\ell$, $m$, the relative displacement amplitude $a_s$, and the inclination $i$ (see Appendix A for a definition of these parameters). To check the stability of our solutions we made use of diagrams, where $\chi^2$ is plotted as a function of every single free parameter. In such diagrams we mark the $\chi^2$-values of all models computed during the optimization. Models having $\chi^2$ below a certain significance limit can be regarded as significant solutions representing the observations.

In a series of four papers, Schrijvers et al. (1997), Telting \& Schrijvers (1997a, 1997b), and Schrijvers \& Telting (1999) have extensively tested the distribution of amplitude and phase across the line profile for different settings of mode input parameters. Therefore, we only focus on the results of fitting the Fourier parameters and in this paper we only present the tests for low-degree modes, where temperature variations are not included in the modeling of the LPV and where the pulsational symmetry axis is aligned with the rotation axis.

\subsection{Low-degree modes} \label{sec:lowdeg}
We tested the quality of our fits for low-degree modes having $0\le \ell \le 4$, all possible positive $m$-values with $0 \le m \le \ell$ and $a_s=0.001$. Our synthetic line profiles were computed by integration over a regular latitude-longitude surface grid consisting of 16200 visible elements, each with a size of $2^{\circ}$x $2^{\circ}$. For each test case we calculated irregularly spaced time series of 300 spectra for a timebase of 12 days with a pulsation frequency of $f=12$~\cd. The projected equatorial rotation velocity was set to \vsini~=~30~\kms; furthermore, we set the intrinsic line width $\sigma=12$~\kms~and the equivalent width $W=8$~\kms. These two values are based on observed values of the \ion{Fe}{i}~$\lambda$5367.467~\AA~line of the $\delta$ Sct star FG Vir. Other physical properties of the model (mass $M=1.85~M_\odot$, radius $R=2.27~R_{\odot}$, effective temperature $T_{\text{eff}}=7250$~K) also resemble those of FG Vir (Breger et al. 1999). We adopted the quadratic limb darkening coefficients $u_{a} = 0.526$ and $u_{b} = 0.163$ listed by Barban et al. (2003) linearly interpolated to the selected wavelength.





To create conditions that are as realistic as possible, the raw synthetic spectra were convolved with a Gaussian instrumental profile resembling a spectrograph with a resolution of R=60 000. Random noise was added to result in S/N=200. 


We derived the uncertainties $\sigma(Z_\lambda)$, $\sigma(A_\lambda)$, and $\sigma(\phi_\lambda)$ from the standard deviation of the residuals across the line profile after prewhitening the input frequency $f$ and its first harmonic $2f$. An estimate of \vsini=30.1~\kms, $\sigma$=11.9~\kms, and $W$=8.0~\kms was derived by fitting the zero point profile. Since the equivalent width $W$ is very well constrained, we set it as a constant throughout the optimization, whereas the other two parameters were set as free in a range of $\pm 0.5$~\kms~of the derived values. In Table~\ref{tab:testlist1}, the three best solutions from a fit to the Fourier parameters are listed for each tested mode setting. 

\begin{table*}[!ht]
\centering
\caption{Mode parameters derived from the application of the FPF method. The following parameter space was searched using genetic optimization (the values  in brackets denote the range [min,max;$\Delta$]): $\ell$~[0,7;1], $m$~[-7,7;1], $a_s$~[0.0001,0.005;0.0001], $i~[5^\circ,90^\circ;5^\circ]$, \vsini~[29.5, 30.5; 0.1]~\kms, $\sigma$~[11.5, 12.5; 0.1]~\kms. For each tested displacement field, the three best solutions are shown. We used the following fixed model input values: pulsational amplitude $a_s=0.001$, stellar inclination angle $i=40^\circ$, projected rotational velocity \vsini=30~\kms, width of the intrinsic profile $\sigma=12$~\kms, and equivalent width $W=8$~\kms. The 95~\%-confidence limit of $\chi^2_\nu$ is 1.35. }
\small
\begin{tabular}{|c|c|c|c|c|c|c||c|c|c|c|c|c|c|c|c|}
\hline
\multicolumn{7}{|c||}{$\ell=0$, $m=0$} & \multicolumn{7}{|c|}{$\ell=1$, $m=0$}\\ \hline
$\chi^2_\nu$ & $\ell$ & $m$ & $a_s$ &$i$ & \vsini & $\sigma$ & $\chi^2_\nu$ & $\ell$ & $m$ & $a_s$ &$i$ & \vsini & $\sigma$\\ \hline
1.31 & 0 & 0 & 0.00103 & 85 & 30.17 & 12.50 &0.99 & 1 & 0 & 0.00103 & 41 & 30.50 & 11.50\\
1.69 & 1 & 0 & 0.00158 & 67 & 30.50 & 12.43 &1.19 & 0 & 0 & 0.00119 & 5 & 29.50 & 11.50\\
8.81 & 2 & 0 & 0.00064 & 15 & 30.43 & 12.03 &6.99 & 2 & 0 & 0.00072 & 5 & 30.30 & 12.10\\
\hline\hline
\multicolumn{7}{|c||}{$\ell=1$, $m=1$} & \multicolumn{7}{|c|}{$\ell=2$, $m=0$}\\ \hline
$\chi^2_\nu$ & $\ell$ & $m$ & $a_s$ &$i$ & \vsini & $\sigma$ & $\chi^2_\nu$ & $\ell$ & $m$ & $a_s$ &$i$ & \vsini & $\sigma$\\ \hline
0.84 & 1 & 1 & 0.00096 & 41 & 29.83 & 11.63 & 1.07 & 2 & 0 & 0.00103 & 41 & 30.37 & 11.97\\
1.02 & 2 & 1 & 0.00080 & 49 & 30.50 & 12.30 & 2.76 & 3 & 0 & 0.00057 & 5 & 30.10 & 12.50\\
5.91 & 2 & 2 & 0.00064 & 67 & 30.30 & 12.03 & 5.81 & 1 & 0 & 0.00057 & 28 & 29.50 & 11.50\\
\hline\hline
\multicolumn{7}{|c||}{$\ell=2$, $m=1$} & \multicolumn{7}{|c|}{$\ell=2$, $m=2$}\\ \hline
$\chi^2_\nu$ & $\ell$ & $m$ & $a_s$ &$i$ & \vsini & $\sigma$ & $\chi^2_\nu$ & $\ell$ & $m$ & $a_s$ &$i$ & \vsini & $\sigma$\\ \hline
0.79 & 2 & 1 & 0.00103 & 39 & 30.10 & 12.50 & 1.17 & 2 & 2 & 0.00096 & 41 & 30.50 & 11.50 \\
0.94 & 1 & 1 & 0.00134 & 36 & 29.50 & 11.90 & 1.33 & 3 & 2 & 0.00072 & 51 & 30.50 & 11.90\\
5.88 & 3 & 1 & 0.00127 & 39 & 30.50 & 12.50 & 3.76 & 3 & 3 & 0.00049 & 82 & 29.97 & 12.17\\
\hline\hline
\multicolumn{7}{|c||}{$\ell=3$, $m=0$} & \multicolumn{7}{|c|}{$\ell=3$, $m=1$}\\ \hline
$\chi^2_\nu$ & $\ell$ & $m$ & $a_s$ &$i$ & \vsini & $\sigma$ & $\chi^2_\nu$ & $\ell$ & $m$ & $a_s$ &$i$ & \vsini & $\sigma$\\ \hline
1.08 & 3 & 0 & 0.00111 & 41 & 29.57 & 12.30 & 0.93 & 3 & 1 & 0.00103 & 41 & 29.57 & 12.37\\
1.12 & 4 & 0 & 0.00080 & 31 & 30.50 & 12.50 & 3.18 & 2 & 1 & 0.00072 & 31 & 29.57 & 11.57\\
2.80 & 5 & 0 & 0.00072 & 23 & 30.50 & 12.30 & 3.60 & 3 & 2 & 0.00103 & 72 & 29.57 & 12.17\\
\hline\hline
\multicolumn{7}{|c||}{$\ell=3$, $m=2$} & \multicolumn{7}{|c|}{$\ell=3$, $m=3$}\\ \hline
$\chi^2_\nu$ & $\ell$ & $m$ & $a_s$ &$i$ & \vsini & $\sigma$ & $\chi^2_\nu$ & $\ell$ & $m$ & $a_s$ &$i$ & \vsini & $\sigma$\\ \hline
0.94 & 3 & 2 & 0.00096 & 46 & 30.43 & 11.50 & 0.92 & 3 & 3 & 0.00088 & 46 & 29.63 & 12.30\\
1.26 & 2 & 2 & 0.00142 & 36 & 30.03 & 11.83 & 1.19 & 4 & 3 & 0.00080 & 54 & 30.50 & 12.50\\
2.64 & 3 & 3 & 0.00072 & 75 & 30.43 & 12.37 & 1.48 & 4 & 2 & 0.00096 & 41 & 30.17 & 12.43\\
\hline\hline
\multicolumn{7}{|c||}{$\ell=4$, $m=0$} & \multicolumn{7}{|c|}{$\ell=4$, $m=1$}\\ \hline
$\chi^2_\nu$ & $\ell$ & $m$ & $a_s$ &$i$ & \vsini & $\sigma$ & $\chi^2_\nu$ & $\ell$ & $m$ & $a_s$ &$i$ & \vsini & $\sigma$\\ \hline
1.13 & 4 & 0 & 0.00088 & 36 & 29.63 & 11.77 & 1.03 & 4 & 1 & 0.00096 & 39 & 29.63 & 12.03\\
1.36 & 5 & 0 & 0.00064 & 18 & 30.17 & 12.03 & 3.03 & 5 & 1 & 0.00096 & 33 & 30.43 & 12.03\\
1.65 & 5 & -2 & 0.00064 & 5 & 30.23 & 11.70 & 3.07 & 4 & 3 & 0.00096 & 13 & 29.97 & 12.23\\
\hline\hline
\multicolumn{7}{|c||}{$\ell=4$, $m=2$} & \multicolumn{7}{|c|}{$\ell=4$, $m=3$}\\ \hline
$\chi^2_\nu$ & $\ell$ & $m$ & $a_s$ &$i$ & \vsini & $\sigma$ & $\chi^2_\nu$ & $\ell$ & $m$ & $a_s$ &$i$ & \vsini & $\sigma$\\ \hline
0.87 & 4 & 2 & 0.00096 & 39 & 30.03 & 11.90 & 1.17 & 4 & 3 & 0.00103 & 41 & 30.03 & 12.30\\
1.04 & 4 & 3 & 0.00088 & 62 & 30.23 & 12.23 & 1.47 & 3 & 3 & 0.00181 & 28 & 30.03 & 11.77\\
1.17 & 3 & 3 & 0.00080 & 49 & 29.50 & 11.63 & 2.63 & 4 & 4 & 0.00103 & 54 & 30.10 & 12.50\\
\hline\hline
\end{tabular}
\begin{tabular}{|c|c|c|c|c|c|c|}
\multicolumn{7}{|c|}{$\ell=4$, $m=4$}\\ \hline
$\chi^2_\nu$ & $\ell$ & $m$ & $a_s$ &$i$ & \vsini & $\sigma$\\ \hline
1.22 & 4 & 4 & 0.00096 & 41 & 30.30 & 11.97\\
1.56 & 5 & 4 & 0.00243 & 85 & 29.97 & 11.50\\
1.62 & 4 & 3 & 0.00088 & 26 & 29.63 & 12.03\\
\hline
\end{tabular}
\vspace*{1cm}
\label{tab:testlist1}
\end{table*}

\begin{figure*}[!ht]
\centering
\begin{tabular}{cc}
\includegraphics*[width=60mm,bb=80 80 464 521,clip,angle=0]{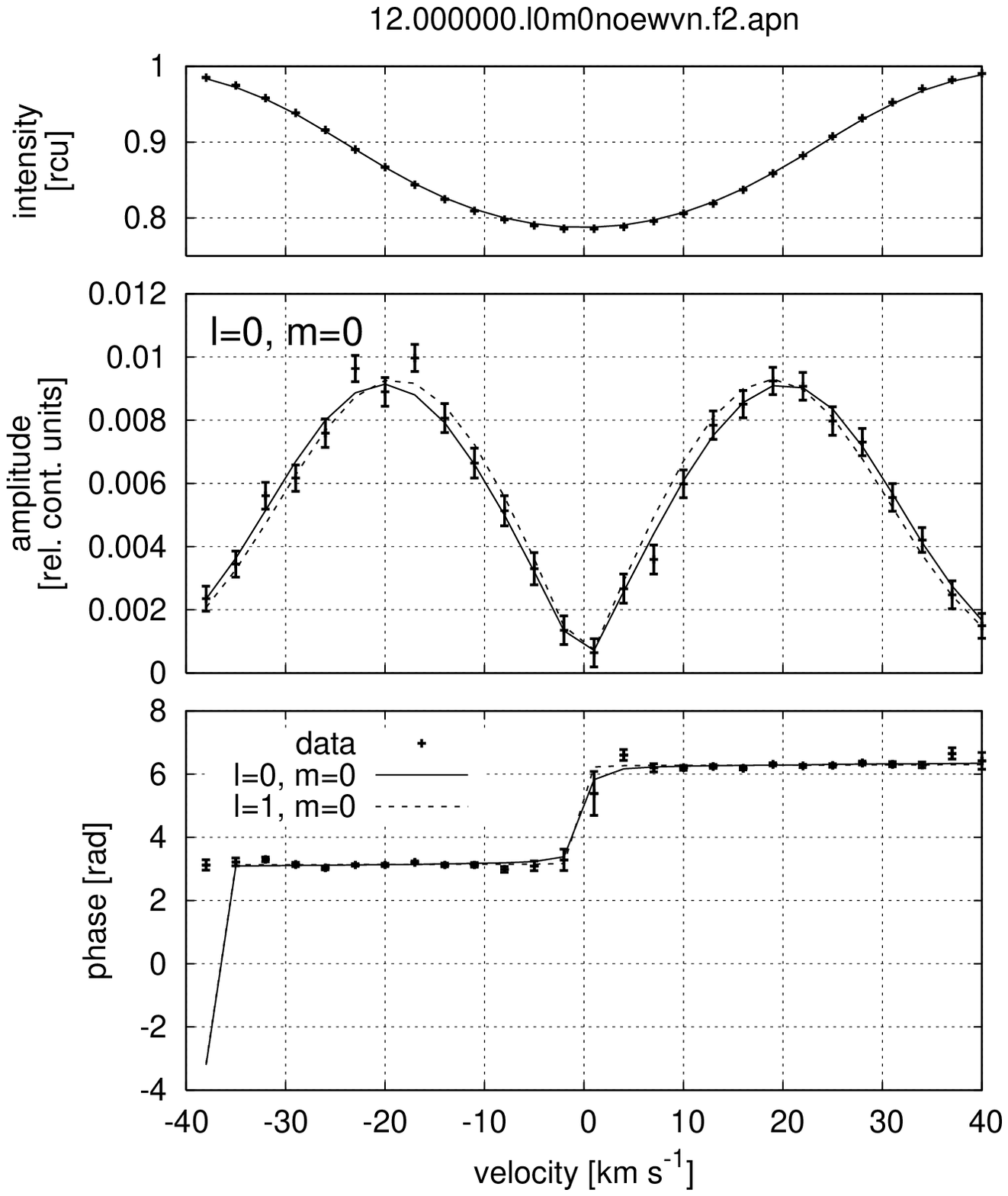}&
\includegraphics*[width=60mm,bb=80 487 277 705,clip,angle=0]{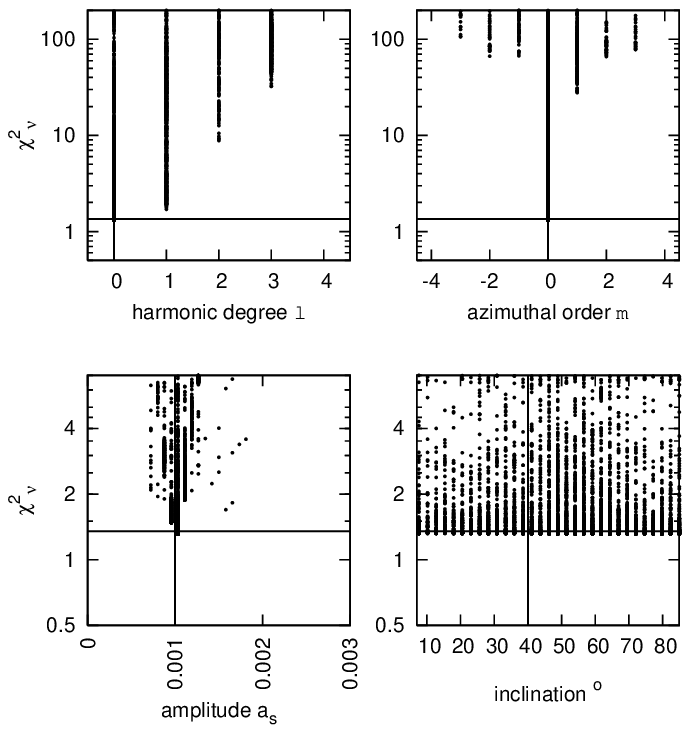}\\
\includegraphics*[width=60mm,bb=80 80 464 521,clip,angle=0]{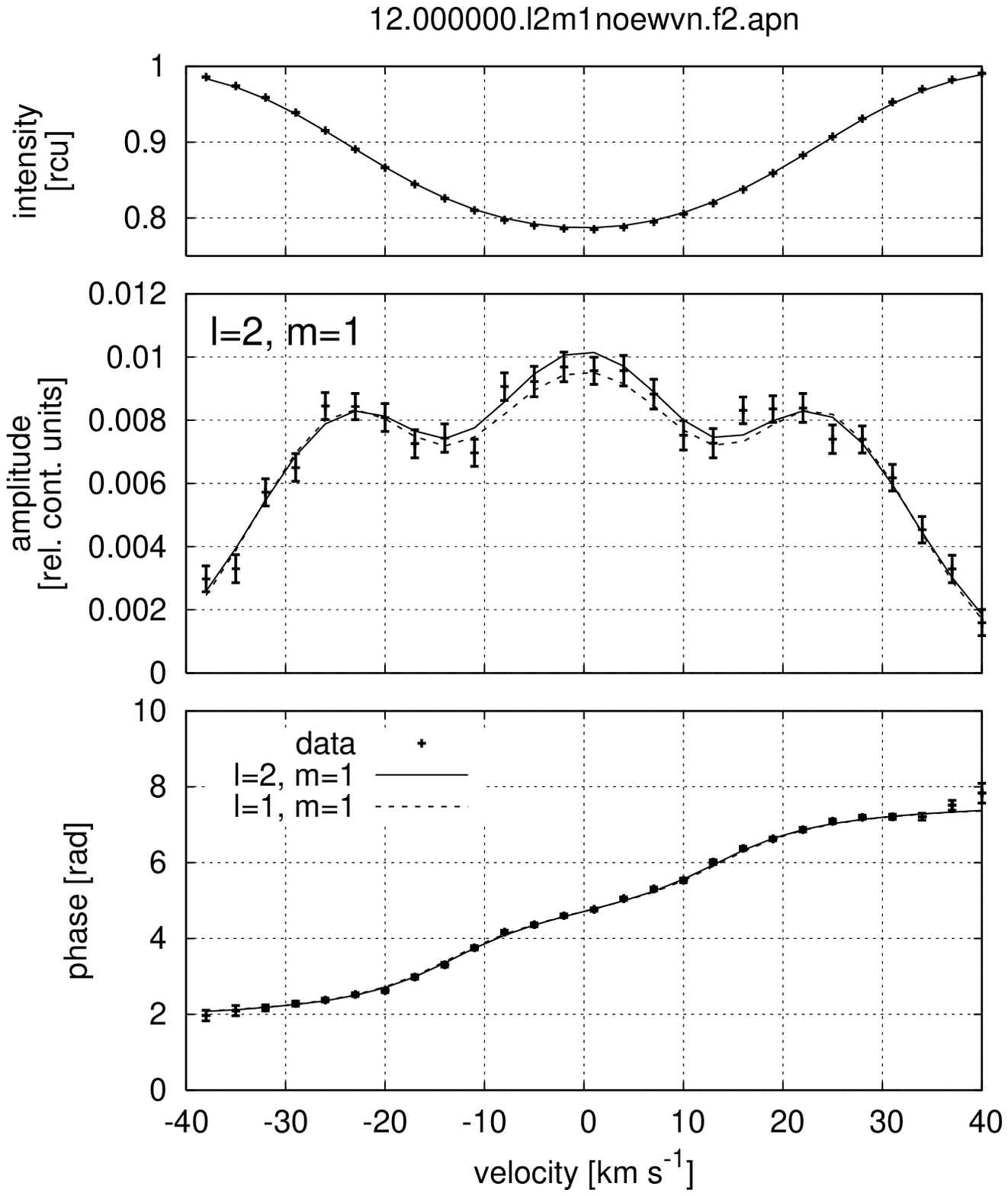}&
\includegraphics*[width=60mm,bb=80 487 277 705,clip,angle=0]{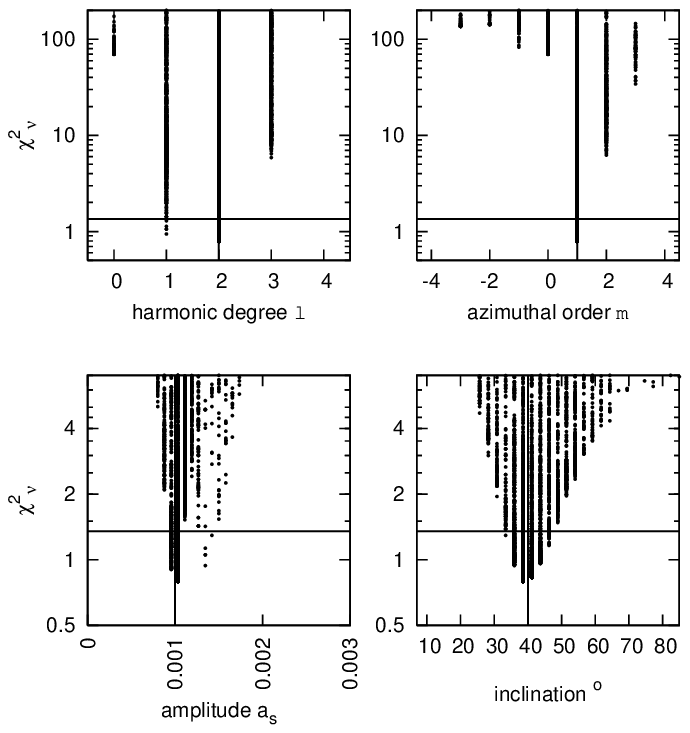}\\
\includegraphics*[width=60mm,bb=80 80 464 521,clip,angle=0]{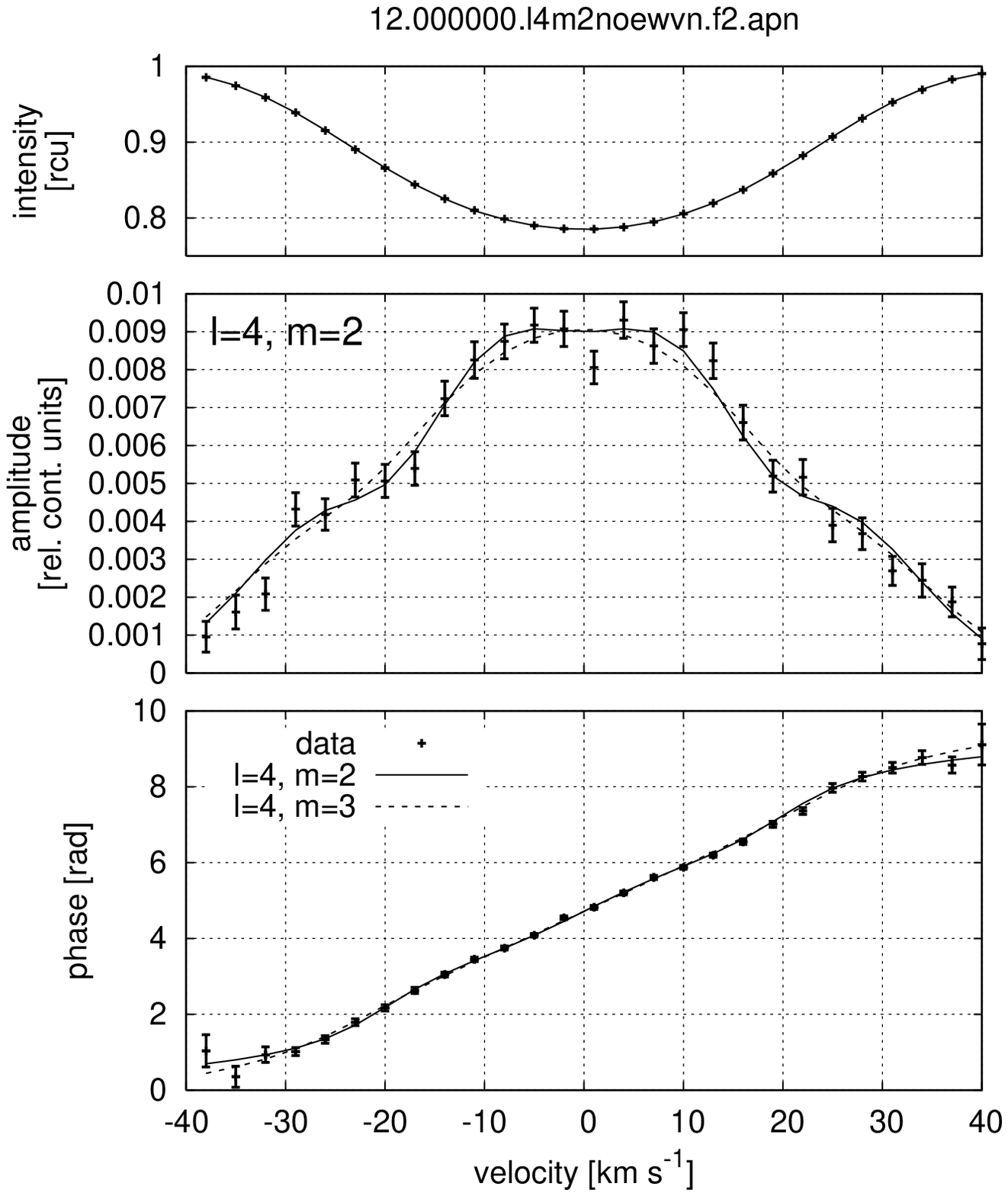}&
\includegraphics*[width=60mm,bb=80 487 277 705,clip,angle=0]{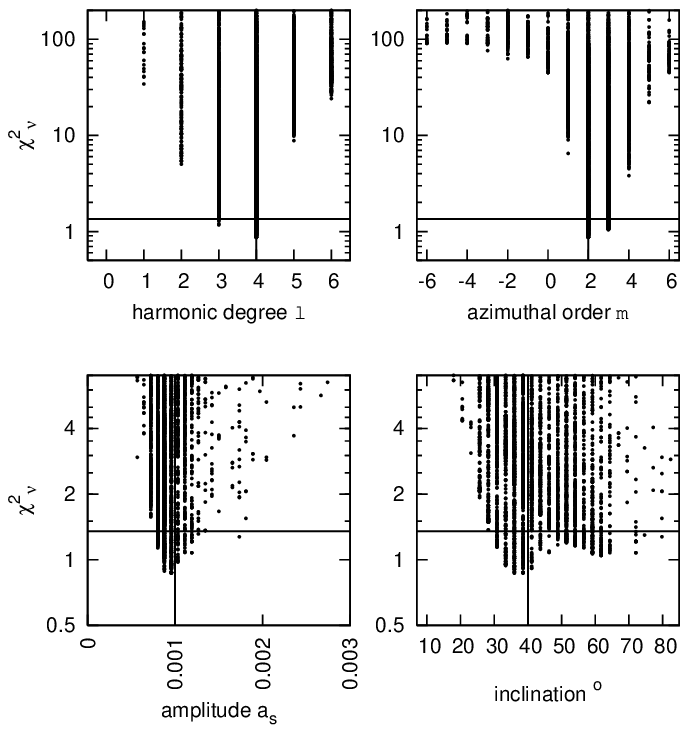}\\
\end{tabular}
\caption{Application of the FPF method to synthetic LPV of three selected displacement fields (300 spectra, S/N=200). The left panels show, from top to bottom, the zero point, amplitude, and phase across the line of the input data (black crosses with error bars) and the two best fits (lines). The four panels on the right hand side show $\chi^2$ with respect to $\ell$ (top left), $m$ (top right), the pulsational amplitude (bottom left), and the inclination angle (bottom left). Every single dot represents a model computed during the genetic optimization. The vertical solid line is at the position of the input parameter values and the horizontal line denotes the 95~\%-confidence limit.}
\label{fig:plot1}
\end{figure*}

For selected modes we present $\chi^2$-diagrams in Fig.~\ref{fig:plot1}, where the behavior of $\chi^2_\nu$ with respect to the free parameters is displayed. Every dot represents a model computed during the optimization. Some general conclusions, which apply to all tested pulsation geometries are the following: 
\vspace*{3mm}
\begin{itemize}
\item For all tested pulsation geometries, the correct identification always had the lowest $\chi^2$. There were sometimes ambiguities where more than one solution was below the 95~\% confidence limit.
\item The value of $m$ is in any case better constrained than $\ell$. The correct identification of $m$ is always in first place, whereas some ambiguity often exists for typing $\ell$. The ambiguities in determinating $\ell$ and $m$ increase for higher degree input modes. 
\item In most cases, the derived values for the amplitude and the inclination are correctly reproduced. For both values a larger uncertainty exists for axisymmetric modes.  There is a small dependence of $\chi^2_\nu$ with respect to the inclination for the $\ell=1, m=0$ - mode, and any inclination angle can be fitted satisfactorily. In contrast, the non-axisymmetric modes show a clear minimum of $\chi^2_\nu$ at or close to the input inclination. The best precision for determining the inclination can be achieved for non-axisymmetric modes with $\ell=1$ or $2$. Identification ambiguities of higher-degree modes increase the uncertainty.

\end{itemize}
We now summarize the conclusions from our tests for different kinds of pulsation modes.
\begin{itemize}
\item{\bf Radial displacement:} Both spherical mode parameters, $\ell$ and $m$, could be derived unambiguously, and, since for the radial mode no dependence of the observed amplitude on the inclination exists, the intrinsic pulsation amplitude could also be derived correctly. When applying lower S/N values to the spectra, it is no longer possible to select between $\ell=0$ and $\ell=1$, whereas only $m=0$ is still below the critical $\chi^2_\nu$ limit.  
\item{\bf Axisymmetric displacement:} The amplitude-phase pattern of these modes is very similar to that of radial pulsation, which often makes a distinction difficult. Again, the value of $m$ can be determined precisely, but ambiguities exist in the typing of $\ell$ and especially in the amplitude and the inclination. The last two parameters directly depend on each other and a wrong inclination can be fitted by tuning the amplitude. 
\item{\bf Tesseral and sectoral displacements} have in common that the determination of the inclination and intrinsic amplitude can be done with higher precision than for axisymmetric modes.
Sectoral modes are more easily seen at higher inclinations, leading to a better identification due to the higher observed amplitude. An interesting aspect is that a misidentification with $\ell_{\text{derived}}=\ell_{\text{input}}+1$ still generally leads to a good estimation of $i$.
\end{itemize}

De Pauw et al. (1993) have carried out extensive numerical tests to explore the accuracy of the MM. Qualitatively, our conclusions for the FPF method agree very well with their results, but we were able to quantify them statistically for the FPF method.

\subsection{Dependence on S/N}
For two displacement fields, $\ell=m=0$ and $\ell=3, m=2$, we tested the effect of different S/N-values on the uncertainty of the mode identification. The stellar and pulsational parameters were identical to those given in Sect.~\ref{sec:lowdeg}, and S/N-levels of 50, 100, 200, 300, 500 and 1000 were tested for a data set consisting of 300 spectra.  

\begin{figure}[!ht]
\centering
\begin{tabular}{cc}
\includegraphics*[width=44mm,bb=77 10 590 541,clip,angle=-90]{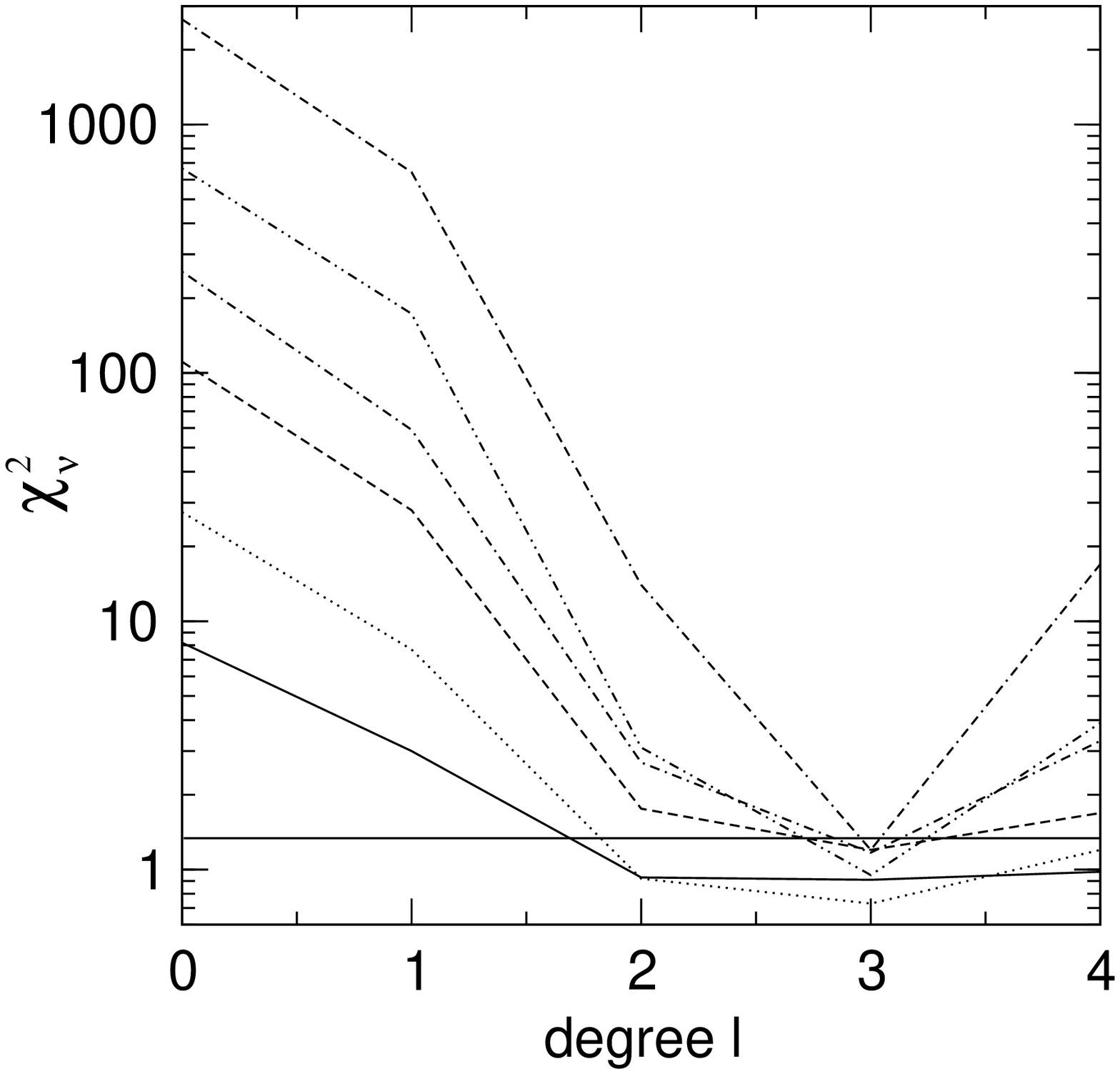}&
\hspace*{-5mm} \includegraphics*[width=44mm,bb=77 70 590 541,clip,angle=-90]{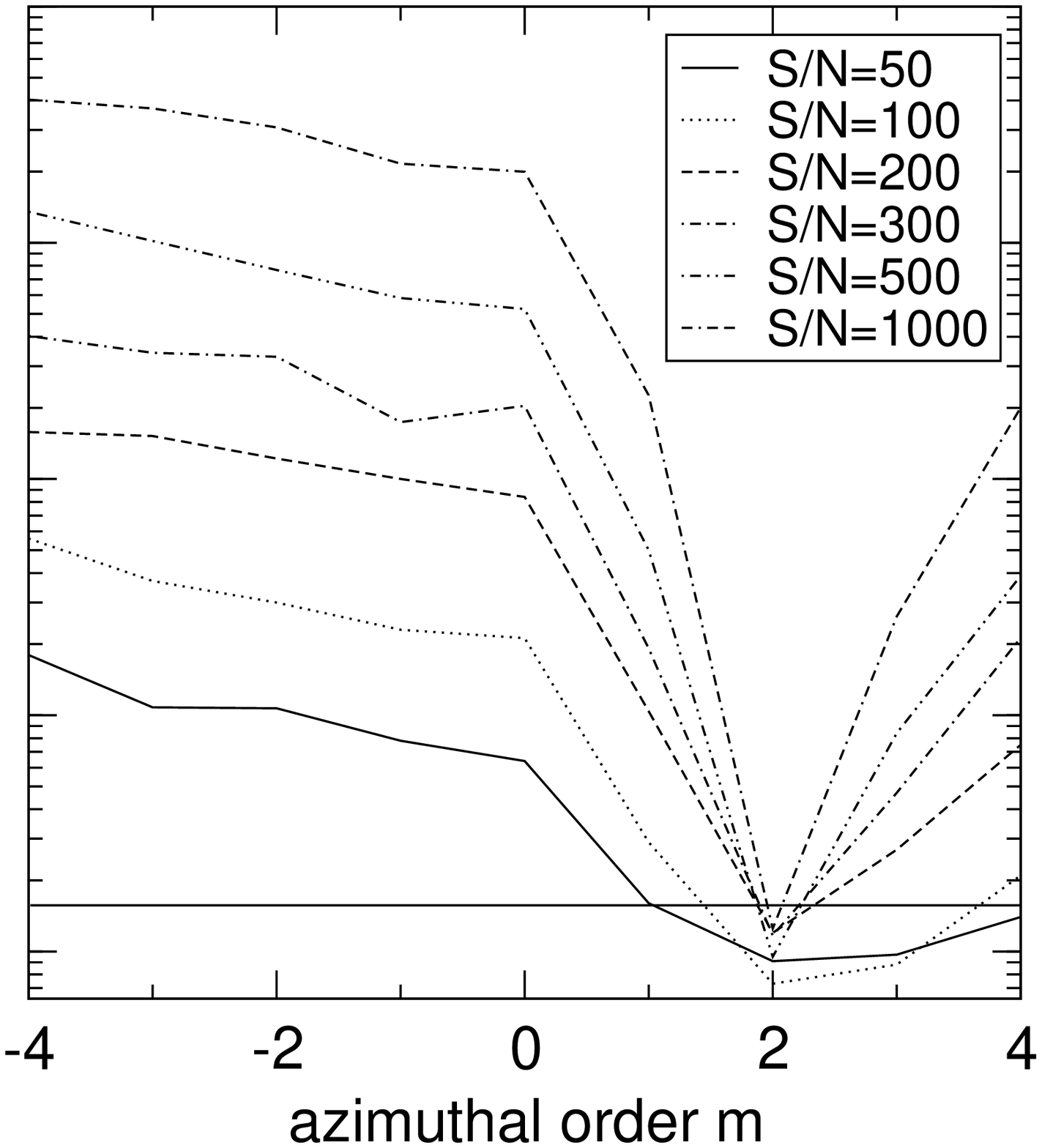}\\
\end{tabular}
\caption{Minimum values of $\chi^2_\nu$ for all tested $\ell$ and $m$ values of an input $\ell=3, m=2$ mode. Every tested S/N is represented by a different line. The solid horizontal line is at the 95~\%-confidence limit. The calculations were made for a fixed number of 300 input spectra.}
\label{fig:sntest}
\end{figure}

As expected, higher S/N ratios lead to less ambiguities in the mode identification. The situation for the tested $\ell=3, m=2$ is reported in Fig.~\ref{fig:sntest}. For the lowest adopted S/N levels, 50 and 100, respectively, no unique identification can be acquired for both $\ell$ and $m$. The situation improves significantly for S/N levels of 200 and higher, where only the fit of the correct input $\ell$ and $m$ values has an $\chi^2_\nu$ value below the 95~\%-confidence limit. These tests show that spectroscopic measurements dedicated to mode identification assuming a number of 300 spectra should at least have an S/N of 200 to be able to provide a well-constrained mode identification.

The variances $\sigma(A_\lambda)^2$ and $\sigma(\phi_\lambda)^2$ of Eqs.~\ref{eq:varr} and \ref{eq:vari} are indirectly proportional to the S/N of the input data and also indirectly proportional to the square root of the number of measurements. Thus, doubling the S/N has the same effect on the mode discrimination as a fourfold number of spectra, i.e., more emphasis should be set on high quality measurements than on the number of spectra. More important than a large data set is the time base of the measurements that determines the frequency resolution, i.e., the ability to separate close frequencies detected in many $\delta$ Scuti stars (Breger \& Pamyatnykh 2006).

\subsection{Multi-mode pulsation}
The experimental results from the mono-mode profile optimization with the Fourier parameter method are also applicable to multi-mode pulsators, if linear pulsation theory applies. The Fourier parameters can be determined by least-square fitting for every single frequency that can be detected from a Fourier analysis of the first moment or of the pixel-by-pixel variations in the line profile. Schrijvers et al. (1997) point out that additional harmonics, beat, and sum frequencies exist in the case of multi-mode pulsation. In our numerical simulations described below, we find that their amplitudes are often too low for detection, especially in the case of $\delta$ Scuti stars. This is also demonstrated for FG Vir by Zima et al. (2006), where among 15 frequencies only one harmonic and one combination term were detected. These additional frequencies cannot be neglected  for objects, such as RR Lyrae stars, showing higher amplitude and must be included in least-square fits to derive correct amplitude and phase values for the intrinsic frequencies.

We tested our method on a five-frequency model consisting of low-degree modes. A similar test has already been successfully carried out for a comparison of the PBP method to the MM by Zima et al. (2004). 

To create conditions that are as realistic as possible, we generated 594 line profiles taking timings from the 2002 FG Vir campaign (Zima et al. 2003), which had a time base of 80 days. We added noise (S/N$~\approx~200$) consisting of the residuals after prewhitening of 15 detected frequencies and simulated the effect of phase smearing for an integration time of 7 minutes. The resulting pulsation pattern of the five pulsation modes was calculated by summing up the contribution of every single spherical harmonic. In these synthetic spectra we did not include temperature effects so as to speed up the process of optimization.

We were able to detect all five input frequencies from a Fourier analysis of the first moment, as well as from the pixel-by-pixel variations, which are more sensitive to higher-degree modes. Due to the relatively high noise, no additional peaks related to combinations or harmonics of the detected frequencies could be found. The mode identification of every single detected mode was carried out in the same way as described for the mono-mode profiles. 

Consistent with the mono-mode fits, again the azimuthal order $m$ can be identified with much higher precision than $\ell$. For all modes $\chi^2_\nu$ of the correct identification of $m$ is below the 95~\%-confidence limit. The precision of the derived $\ell$ is about $\pm~1$ for the tested modes. 
\begin{figure}[!ht]
\centering
\includegraphics[width=64mm,bb=78 22 577 717,clip,angle=-90]{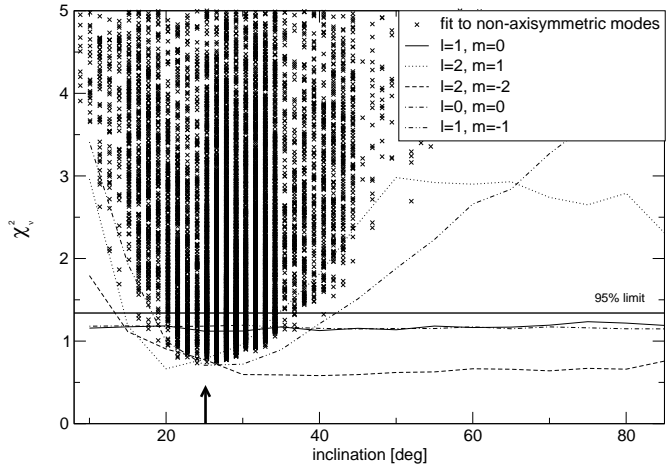}\\
\caption{Determination of the stellar inclination angle from the application of the FPF method to synthetic line profiles ($v \sin i=30$~km~s$^{-1}$, $i=25^\circ$). The derived $\chi^2_\nu$-values are shown as a function of the inclination for different displacement fields (solid lines) and of a multi-mode fit to the non-axisymmetric modes (crosses), respectively. Axisymmetric modes are not well-suited to deriving the inclination angle. A simultaneous fit to all non-axisymmetric modes provides the best estimate of the inclination.}
\label{fig:incli}
\end{figure}

An important result is that the stellar inclination angle and henceforth also the modes' intrinsic pulsation amplitude is better constrained for the non-axisymmetric modes (see Fig.~\ref{fig:incli}). For axisymmetric modes, the sensitivity of $\chi^2$ with respect to the inclination is very small due to the fact that we cannot decouple the intrinsic amplitude and the inclination and, since we do not know the intrinsic amplitude, we can scale it such that any inclination can be fitted satisfactorily. Especially the fact that an axisymmetric mode might be a radial mode makes it necessary to judge inclination values derived from such a mode with great caution.

An improved approach to deriving a more precise inclination value can be carried out by a simultaneous fit to the Fourier parameters of all detected non-axisymmetric modes with common values of \vsini, $\sigma$, and $W$. By doing so, we take advantage of multi-periodic pulsation, assuming that all observed modes are seen at the same aspect angle. Figure~\ref{fig:incli} shows the comparison of the derived $\chi^2_\nu$ values for the five input modes as a function of the inclination. Whereas no significant minimum exists for the two axisymmetric modes $f_1$ and $f_4$, there is a clear dependence on $\chi^2_\nu$ and the inclination for the other modes. The simultaneous fit of the three non-axisymmetric modes yields an inclination value of $25\pm10^\circ$, which is consistent with the input value.

\subsection{Comparison with the MM and the PPM}
We carried out a comparison of the FPF method with two other spectroscopic mode identification methods: the MM and the PPM. This comparison was accomplished by analyzing the same synthetic data set with all three methods. We used the data of the comparison test described in Zima et al. (2004), which consists of a time series of synthetic line profiles distorted by a 6-frequency displacement field. The input modes have a low degree ($0 \le \ell \le 3$), and the profiles are computed for a star having a low \vsini~of 21~\kms.
\begin{figure*}[!ht]
\centering
\includegraphics[width=120mm,bb=12 8 318 121,clip,angle=0]{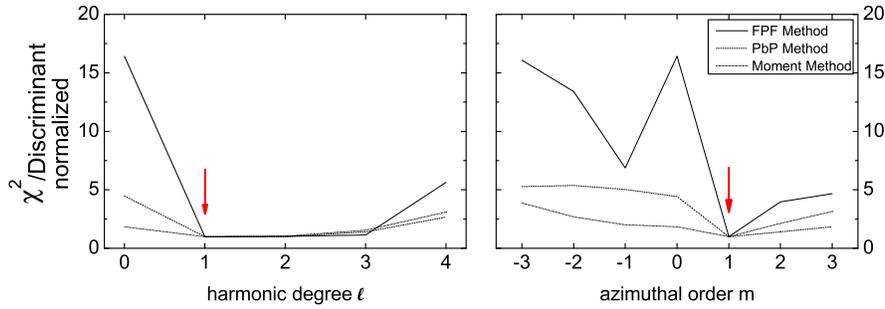}\\
\caption{Comparison of the identification of $\ell$ and $m$ derived from applying the FPF, PPM, and MM to synthetic line profiles. The best derived deviation parameters for each tested value of $\ell$ and $m$ is shown for the different methods. For comparison, the deviation values have been normalized. The model input values $\ell=m=1$ are marked by the arrows.}
\label{fig:comp}
\end{figure*}
Since the three methods quantify the quality of their fits, which we call deviation parameter, in mathematically different ways, the obtained values were normalized to the minimum derived value to permit a meaningful comparison. 

The striking difference between the methods lies in the sensitivity of the deviation parameters with respect to different solutions. This is illustrated in Fig.~\ref{fig:comp} where the computed deviation parameters of the three methods are shown for one of the identified displacement fields ($\ell=m=1$). The minimum deviation value of the FPF method is more pronounced, especially for the azimuthal order $m$, than the minimum of the two other methods, thus implying an improved constraint on the mode identification. The situation is similar for the other five identified displacement fields. The value of $m$ is better constrained for every test case than the harmonic degree $\ell$.

The higher sensitivity of the FPF method compared to the two other methods is mainly due to the fact that only the deviations from the mean profile are taken into account for the calculation of $\chi^2$. Therefore, the fitting puts emphasis on the time variable parameters and consequently on the determination of $\ell$, $m$, $a_s$, and $i$, whereas less emphasis is put on \vsini, $\sigma$, and $W$.

We conclude that by means of the FPF method, especially $m$ can be determined with much better precision than with the MM and the PPM. The high sensitivity of $\chi^2$ with respect to the azimuthal order increases the probability that only one $m$-value lies below the 95~\%-confidence limit. In our comparison we made the assumption that systematic errors in the continuum normalization do not occur, which is often not true for real data. In fact the FPF method and the PPM are much more affected by normalization uncertainties than the MM. Any such systematic uncertainties would reduce the sensitivity of the first two methods compared to the MM. The behavior of the deviation parameters with respect to $\ell$ is quite similar for all three tested methods, leading to ambiguities in the identification of this parameter.

\section{Application limits}
We examined the limits for applying the FPF method in terms of projected rotational velocity and radial velocity amplitude. A major limit is set by the pulsation amplitude of the modes especially in a multi-periodically pulsating star. If the ratio of the projected radial velocity amplitude (RVA) and the equivalent width of at least one mode is too high, other modes are no longer projected linearly onto the Doppler velocity scale relative to the center of the stellar disk. Furthermore, a significant fraction of the power of the modes is transferred to their combination terms decreasing the power of the intrinsic frequencies. The Fourier values of such modes are distorted and cannot be represented by a single-mode model. The zero point profile is broadened and will lead to an overestimation of \vsini.
\begin{figure*}[!ht]
\centering
\begin{center}
 \begin{tabular}{cc}
\resizebox{60mm}{!}{\includegraphics*[width=40mm,height=36mm,bb=9 24 157 145,clip,angle=0]{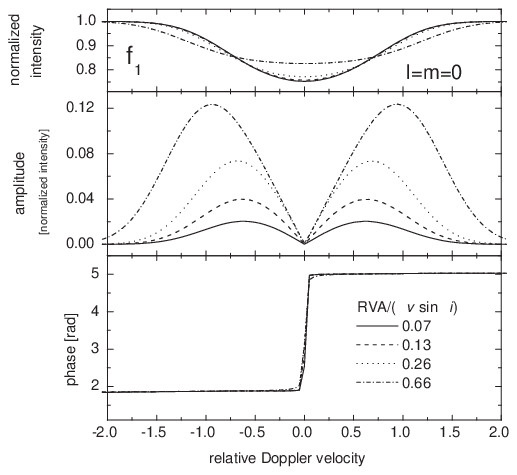}} &
\resizebox{60mm}{!}{\includegraphics*[width=40mm,height=36mm,bb=11 24 160 145,clip,angle=0]{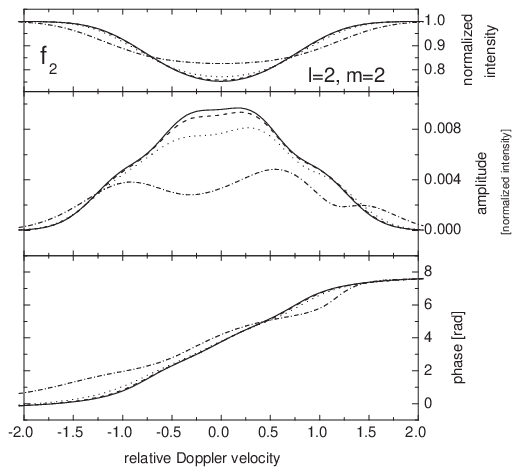}}\\
\resizebox{60mm}{!}{\includegraphics*[width=40mm,height=40mm,bb=9 11 160 145,clip,angle=0]{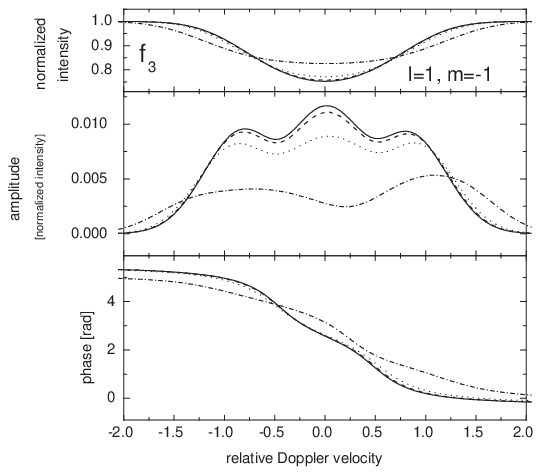}} &
\resizebox{60mm}{!}{\includegraphics*[width=40mm,height=40mm,bb=11 11 160 145,clip,angle=0]{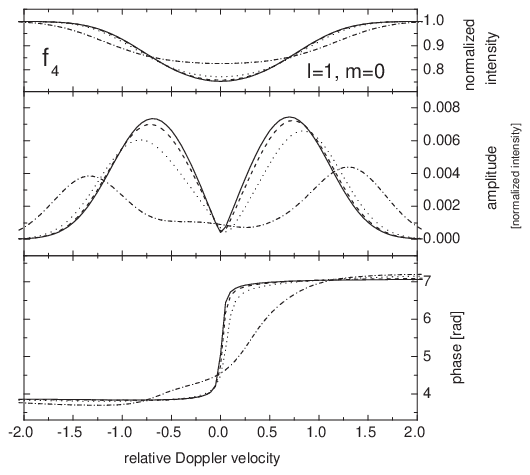}}\\
\end{tabular}
\end{center}
\caption{Influence of a displacement field having a large radial velocity amplitude (RVA) on the amplitude and phase of other simultaneously excited pulsation modes. The data were calculated from an experimental four-frequency multi-mode model. The computations were made for different amplitudes of $f_1$, whereas the amplitudes of the other modes were not modified. The abscissa represents the Doppler velocity normalized to \vsini.}
\label{fig:large1}
\end{figure*}%

This limit applies especially to narrow-lined stars with relatively large amplitudes, such as RR Lyr (Kolenberg 2002) or $\nu$ Eri (De Ridder et al. 2004), where spectroscopic mode identification using Fourier values contradicts the results obtained from photometry. For these cases the MM is much better suited since it uses integrated flux across the line.

We tested this problem for a four-frequency multi-mode model and calculated Fourier parameter diagrams for different intrinsic pulsation amplitudes of $f_1$, keeping the amplitudes of the other modes constant. The resulting diagrams are reported in Fig.~\ref{fig:large1}. A significant reduction of the amplitude across the profile occurs for $f_2$, $f_3$, and $f_4$ for (RVA/\vsini)$_{f_1}$~above 0.13. 

\begin{table}[!ht]
\centering
\caption{Overestimation of \vsini~due to pulsational broadening of the zero point profile by a radial displacement field. The subscript $i$ denotes input values, whereas $d$ denotes derived values.}
\begin{tabular}{c|c}
RVA/\vsini & $\frac{[v \sin i]_i}{[v \sin i]_d}$ \\ \hline
0.003 & 1.000\\
0.016 & 1.000\\
0.032 & 1.001\\ 
0.065 & 1.007\\
0.100 & 1.010\\
0.163 & 1.026\\
0.327 & 1.131\\
0.653 & 1.455\\
\end{tabular}

\label{tab:zerobroadening}
\end{table}
The average and zero point profile respectively are broadened by a large amplitude mode. The effect of the overestimation of \vsini~due to the pulsationally caused broadening was explored by fitting the zero point profile with synthetic counterparts. The results are reported in Table~\ref{tab:zerobroadening}. For different arbitrary values of RVA/\vsini, the ratio of the input to the derived \vsini~is given. Up to an RVA that is 10\% of \vsini, the error of the derived \vsini~is below 1\%. For ratios above 30\%, where an error larger than 10\% is induced, it is better to derive \vsini~by fitting the narrowest available line profile.

Other mode identification methods that rely on the calculation of Fourier parameters, such as the PPM or the IPS method, are affected in the same way. For stars with large RVA relative to their rotational line broadening, these methods have to be applied with great caution. The MM is much better suited to the mode identification of such objects.

\section{Conclusions}
We have presented a new spectroscopic mode identification method, the FPF method, based on fitting the Fourier parameters of the intensity variations across the line profile. Our description of the displacement field is valid for a slowly rotating non-radially pulsating star and includes toroidal motion, i.e., first-order effects due to the Coriolis force. The implementation allows a modeling of pulsational geometries where the pulsational symmetry axis is not aligned with the stellar rotation axis. Temperature variations of the stellar atmosphere and their impact on the equivalent width variations of a line can be taken into account.

The strength of the FPF method is its ability to estimate the significance of the derived mode parameters by means of an $\chi^2$ test. For a given error threshold it is possible to exclude a number of tested $\ell$ and $m$ values for a pulsation mode. Furthermore, the stability of the derived fits can be determined by analyzing the trend of the $\chi^2$ values across the tested parameter ranges. Compared to previous methods, this permits a statistical quantification of the obtained solutions. We have also shown that the FPF method can determine $m$ with better precision than other methods. This is especially important for improving the treatment of rotational effects in theoretical seismic models (Pamyatnykh 2003).
   

The presented method is optimized for slowly to intermediately rotating pulsating stars that have a large projected rotational velocity compared to the pulsational radial velocity amplitude. Such objects can be found especially among $\delta$ Scuti and $\beta$ Cephei stars (Aerts \& De Cat 2003). In a subsequent paper we demonstrate the successful application of the FPF method to high-dispersion time-series spectra of the non-radial pulsating $\delta$ Scuti star FG Vir (HD 106384).

\appendix
\section{Computation of the distorted line profile and flux variations}
\subsection{The displacement and velocity field}
We assume an unperturbed model star to be spherically symmetric, in hydrostatic equilibrium, and in the absence of a magnetic field and rotation. The position of a mass element of such a star can be written in spherical coordinates $(r,\theta,\phi)$ defined by the radius $r$, the colatitude $\theta~[0,\pi]$, i.e., the angular distance from the pole, and the azimuth angle $\phi~[0,2\pi]$. Any shift of a mass element from its equilibrium position is given by the Lagrangian displacement vector $\pmb{\xi}=(\xi_r,\xi_\theta,\xi_\phi)$. This displacement modifies the initial pressure $p_0$, the density $\varrho_0$, and the gravitational potential $\Phi_0$ as a function of $r$, $\theta$, $\phi$, and the time $t$. The linear, adiabatic perturbations of these parameters are governed by the four equations of hydrodynamics, i.e., Poisson's equation, the equation of motion, the equation of continuity, and the condition for adiabacity.

This set of differential equations is solved by assuming that $r$, $\theta$, and $\phi$ depend on $Y_\ell^m(\theta,\phi)~e^{i\omega t}$, where $Y_\ell^m(\theta,\phi)$ (Abramowitz \& Stegun 1964) denotes the spherical harmonic of degree $\ell$ and of order $m$, $\omega$ is the angular pulsation frequency, and $t$ the time. The spherical harmonic can be written as 
\begin{equation}
Y^m_\ell (\theta,\phi) \equiv N^m_\ell P^{|m|}_\ell (\cos \theta)~e^{im\phi}.
\label{eq:ylm}
\end{equation}
Here, $P_\ell^{|m|}$ denotes the associated Legendre function of degree $\ell$ and azimuthal order $m$ given by
\begin{equation}
P^m_\ell(x) \equiv \frac{(-1)^m}{2^\ell \ell!} (1-x^2)^{\frac{m}{2}} \frac{d^{\ell+m}}{dx^{\ell+m}}(x^2-1)^\ell,
\end{equation}
and
\begin{equation}
N^m_\ell = (-1)^\frac{m+|m|}{2}\sqrt{\frac{(2\ell+1)}{4\pi} \frac{(\ell-|m|)!}{(\ell+|m|)!}}
\end{equation}
is a normalization constant.
The definition of $N^m_\ell$ changes from author to author, which must be taken into account when comparing derived amplitudes.

We model uniform stellar rotation, including first-order corrections due to the Coriolis force, which gives rise to toroidal motion. The resulting displacement field in the case of a slowly rotating non-radially pulsating star cannot be described by a single spherical harmonic anymore. It consists of one spheroidal and two toroidal terms, which only have a horizontal component, and is given for an angular frequency $\omega$ in the stellar frame of reference and a time $t$ by
\begin{equation}
\begin{split}
\pmb{\xi} &= \sqrt{4\pi}\biggl[a_{s,\ell}\Biggl(1,k\frac{\partial}{\partial\theta},k\frac{1}{\sin\theta}\frac{\partial}{\partial\phi}\Biggr)~Y^m_\ell(\theta,\phi)~e^{i\omega t}\\
&+ a_{t,\ell+1}\Biggl(0,\frac{1}{\sin\theta}\frac{\partial}{\partial\phi},-\frac{\partial}{\partial\theta},\Biggr)~Y^m_{\ell+1}(\theta,\phi)~e^{i(\omega t+\frac{\pi}{2})}\\
&+ a_{t,\ell-1}\Biggl(0,\frac{1}{\sin\theta}\frac{\partial}{\partial\phi},-\frac{\partial}{\partial\theta},\Biggr)~Y^m_{\ell-1}(\theta,\phi)~e^{i(\omega t-\frac{\pi}{2})}\biggr]\\
\end{split}
\label{eq:xi}
\end{equation}
(Martens \& Smeyers 1982, Aerts \& Waelkens 1993, Schrijvers et al. 1997). Note that the term proportional to $Y^m_{\ell-1}$ is not defined for radial and sectoral modes.
Here, $a_{s,\ell}$ denotes the amplitude of the spheroidal component of the displacement field, whereas $a_{t,\ell+1}$ and $a_{t,\ell-1}$ are the corresponding amplitudes of the toroidal components. We neglect the first order correction of the amplitude $a_s$ due to rotation, whereby the amplitudes of the toroidal terms can be approximated by the following relations
\begin{equation}
\begin{split}
a_{t,\ell+1} &= a_{s,\ell}~\frac{\Omega}{\omega} \frac{\ell-|m|+1}{\ell+1} \frac{2}{2\ell+1} (1-\ell k)\\
a_{t,\ell-1} &= a_{s,\ell}~\frac{\Omega}{\omega} \frac{\ell+|m|}{\ell} \frac{2}{2\ell+1} \biggl(1+(\ell+1) k\biggr).\\
\end{split}
\end{equation}
 The factor $\sqrt{4\pi}$ in Eq.~\ref{eq:xi} is introduced in order to scale the normalization $\sqrt{4\pi} N_0^0=1$, such that $a_s$ represents the fractional radius variation for radial pulsation.

The ratio of the horizontal to vertical amplitude, which allows the distinction between $p$- and $g$-modes, can be approximated in the limit of slow rotation by the following relation (Schrijvers et al. 1997) 
\begin{equation}
k \equiv \frac{a_h}{a_s} = k_0+2m\frac{\Omega}{\omega}\biggl[\frac{1+k_0}{\ell(\ell+1)}-C_{\ell,n}\biggr]
\label{eq:k}
\end{equation}
where $a_h$ and $a_s$ are the horizontal and vertical amplitude and $C_{\ell,n}$ is the Ledoux constant. Likewise, $k_0=\frac{GM}{\omega^2 R^3}$ denotes the ratio of the horizontal to vertical amplitude in the limit of no rotation. Here, $M$ and $R$ are the stellar mass and radius, and $G$ is the gravitational constant. 

The displaced surface $\pmb{S}=(S_r,S_\theta,S_\phi)$ is calculated from the sum of the undisplaced coordinates of the unit sphere and the displacement vector by
\begin{equation}
\begin{pmatrix}S_r\\S_\theta\\S_\phi\end{pmatrix} \equiv \begin{pmatrix}1+\xi_r\\ \theta+\xi_\theta\\ \phi+\xi_\phi+\phi_o\end{pmatrix}. 
\end{equation}
Here, the temporal dependence of the displacement field is already incorporated in the definition of $\pmb{\xi}$ from Eq.~\ref{eq:xi}. If the pulsation axis is tilted by an angle of $j$ with respect to the stellar rotation axis, the resulting displaced surface is shifted by a transformation of the spherical coordinates by 
\begin{equation}
\begin{split}
\cos \theta_j &= \cos \theta \cos j + \sin \theta \cos \phi \sin j\\
\sin \theta_j &= \sqrt{1-\cos^2 \theta_j}\\
\cos \phi_j &= (\sin \theta \cos \phi \cos j - \cos \theta \sin j)/\sin \theta_j\\
\sin \phi_j &= \sin \phi \sin \theta / \sin \theta_j.\\
\end{split}
\end{equation}

In our implementation we divide a sphere of unit radius into a grid of regular segments each having equal dimensions in $\Delta\theta$ and $\Delta\phi$. The displaced coordinates of the surface $\pmb{S}$ are evaluated for every single point of the grid. 

The rotation of the surface is implemented by moving the observer around the surface, which is fixed at the origin of the coordinate system. At time $t$ and the inclination angle $i$, the spherical coordinates of the observer then are $\pmb{O}=(1,i,\Omega t)$.  By orthographic projection, we convert the three-dimensional surface coordinates to a plane normal to the line of sight to two-dimensional coordinates by
\begin{equation}
\begin{pmatrix}x\\y\end{pmatrix}=S_r
\begin{pmatrix}
\sin S_\theta \sin (S_\phi-\Omega t)\\
\cos S_\theta \sin i-\sin S_\theta \cos i \cos (S_\phi-\Omega t)\\
\end{pmatrix}.
\end{equation}
These coordinates are used to calculate the area $A(\theta,\phi,z)$ of the projected surface segments and for displaying the distorted surface on a screen. At each time step of the calculations, the visible part of the surface $\pmb{S}$ is determined by computing the angle $z$ between the line of sight and the numerically derived surface normal of every single segment. Visibility is provided if $0 \le z \le \frac{\pi}{2}$. Deviations from spherical symmetry due to the surface displacement field are taken into account for the calculation of $z$. 

We compute the velocity field of the displaced surface from the numerical temporal derivative of the surface coordinates. The advantage over an analytical approach is its higher flexibility if the description of the displacement field is modified, e.g., if additional rotational effects are incorporated. The velocity component $V$ in the direction of the observer finally is given by
\begin{equation}
V = V_x \sin i \cos (\Omega t)+V_y \sin i \sin (\Omega t) +V_z \cos i,
\label{eq:V}
\end{equation}
where $(V_x,V_y,V_z)$ are the Cartesian coordinates of the derived surface velocity field.

\subsection{Construction of the distorted line profile}
We assume that the intrinsic line profile is a Gaussian, which may undergo equivalent width changes due to temperature variations. 
The distorted line profile is calculated from an integration of an intrinsic profile over the whole visible stellar surface, which - for computational purposes - numerically results in a weighted summation over the surface grid.
We define the intrinsic Gaussian profile in a surface point having the line-of-sight velocity $V$ (see Eq.~\ref{eq:V}) as
\begin{equation}
I(v,T_{\text{eff}},\logg)=\biggl(1+\frac{\delta F}{F}\biggr)\biggl[1-\frac{W_\text{int}(T_{\text{eff}})}{\sigma\sqrt{\pi}}e^{-(\frac{V-v}{\sigma})^2}\biggr].
\end{equation}
Here, $v$ is the velocity across the line profile, $\delta F/F$, which will be defined in Section~\ref{sec:flux}, takes the surface flux of the emitting segment into account, $W_\text{int}(T_\text{eff})$ is the equivalent width as a function of the effective temperature (see Eq.~\ref{eq:ewvariation}); and $\sigma$ is the width of the intrinsic profile. Note that in our definition the value of $\sigma$ is higher by a factor of $\sqrt{2}$ compared to the definition by Aerts et al. (1992) generally adopted for the MM. 

The distorted profile $\frak{I}$ can be calculated by summation over all visible segments on the surface grid of ($\theta,\phi$) weighted over the projected surface and adopting a quadratic limb darkening law by
\begin{equation}
\begin{split}
\frak{I}&=\sum_{\theta,\phi}\Biggl[\delta(\theta,\phi,z)  A(\theta,\phi,z) \biggl(1-[u_a(1-\mu)+u_b(1-\mu)^2]\biggr)\\
&\phantom{=\sum_{\theta,\phi}\Biggl[}\sum_{v}I(v,T_{\text{eff}},\logg)\Biggr].\\
\end{split}
\label{eq:profile}
\end{equation}
Here, $\delta(\theta,\phi,z)=1$ if the surface element at ($\theta,\phi$) is visible and $\delta(\theta,\phi,z)=0$ otherwise. Also, $\mu=\cos z$ is the cosine of the angle between the surface normal and the line of sight, $u_a$ and $u_b$ are the first and second order limb darkening coefficients, and $A(\theta,\phi,z)$ is the projected surface area of the surface element.

The response of a line's equivalent width to local temperature changes is dependent on the involved element, its excitation, and the temperature in the zone where the line originates. In order to take this effect into account, a variable equivalent width of the intrinsic line profiles must be considered for calculating the distorted profile. Since there is no phase shift between $\delta W_{\text{int}}(T)$ and $\delta T$ we can write, following Schrijvers \& Telting (1999),
\begin{equation}
W_{\text{int}}(T_{\text{eff}})=W_0 (1+\alpha_\text{W} \delta T_{\text{eff}}),
\label{eq:ewvariation}
\end{equation}
where $\alpha_\text{W}$ is a parameter denoting the equivalent width's linear dependence on $\delta T_\text{eff}$, which can be approximated for $\delta T_\text{eff} \ll 1$. A description of the temperature variations will be provided in the next section.

\subsection{Flux variations} \label{sec:flux}
For calculating the local temperature, surface gravity, and flux variations, we closely followed Balona (2000) and Daszynska-Daszkiewicz et al. (2002). Since the flux variation $\delta F/F$ is mainly a function of $T_{\text{eff}}$ and $\log g$, we can write in the limit of linear pulsation theory
\begin{equation}
\begin{split}
\frac{\delta F}{F} &= \alpha_T \frac{\delta T_{\text{eff}}}{T_{\text{eff}}} + \alpha_g \frac{\delta g}{g}=\\
&= \frac{\delta R}{R_0}\biggl[\alpha_T f \frac{1}{4} e^{i\psi_f} - \alpha_g \biggl(2+\frac{3\omega^2}{4\pi G <\rho>}\biggr)\biggr],\\
\label{eq:dF}
\end{split}
\end{equation}
 where $\alpha_T$ and $\alpha_g$ given by
\begin{equation}
\alpha_T=\biggl(\frac{\partial \log F}{\partial \log T_{\text{eff}}}\biggl)_g~\text{and}~\alpha_g=\biggl(\frac{\partial \log F}{\partial \log g}\biggl)_{T_{\text{eff}}} 
\end{equation}
are partial derivatives of the flux, which can be calculated from static model atmospheres for different passbands.

 Here, $R_0$ is the unperturbed radius, $G$ denotes the gravitational constant, $<\rho>$ is the mean density of the star, $f$ the absolute value of the complex $f_R+if_I$, and $\psi_f$ the phase lag of the displacement between the radius and temperature eigenfunctions. Then $f$ describes the ratio of flux to radius variations, which can be transformed into the ratio of temperature to radius variations due to the fact that the flux is proportional to $T^4$.

The total flux $\frak{F}$ is derived by weighted summation over the visible surface
\begin{equation}
\begin{split}
\frak{F}&=\sum_{\theta,\phi}\Biggl[\delta(\theta,\phi,z) A(\theta,\phi,z) \biggl(1-[u_a(1-\mu)+u_b(1-\mu)^2]\biggr)\\
&\phantom{=\sum_{\theta,\phi}\Biggl[}\biggl(1+\frac{\delta F}{F}\biggr)\Biggr]\\
\end{split}
\label{eq:flux}
\end{equation}
where the symbols have the same meaning as in Eq.~\ref{eq:profile}.

By definition, $f$ reflects the temperature change due to the radius variations during pulsation. Then $\delta T_\text{eff}$ in turn governs the equivalent width variations of a line and the flux variations. Therefore, we can determine empirical values of $f$ by two independent approaches. The derived values of $f$ yield information about sub-photospheric layers where the thermal time scale is on the order of the pulsation period. These layers are only poorly probed by seismic modeling of only frequency values.

\begin{acknowledgements}
This work was supported by the Austrian Fonds zur F\"orderung der wissenschaftlichen Forschung (Project P17441-N02). The author wants to thank Michel Breger, Katrien Kolenberg and Jagoda Daszynska-Daszkiewicz for important discussions.
\end{acknowledgements}

\end{document}